%% file: main.tex
\documentclass[manuscript]{aastex}
\usepackage{amsmath}
\usepackage{epsfig}

\input definitions.tex

\input Tables.tex

\input Figures.tex

\shorttitle{High-Rayleigh-Number Stellar Convection}
\shortauthors{Featherstone \& Hindman}

\begin{document}

\title{The Spectral Amplitude of Stellar Convection and its Scaling in the High-Rayleigh-Number Regime}

\author{Nicholas A. Featherstone}
\affil{Department of Applied Mathematics, University of Colorado, Boulder, CO 80309-0526}
\email{feathern@colorado.edu}
\author{Bradley W. Hindman}
\affil{JILA \& Department of Astrophysical and Planetary Sciences, University of Colorado, Boulder, CO 80309-0440}

\input Abstract.tex


\input Introduction.tex

\input Model.tex

\input Results.tex

\input Conclusions.tex

\acknowledgments
We wish to thank Mark Miesch and Juri Toomre for several useful discussions over the years that helped to frame the central question posed in this work.  We further wish to thank Keith Julien for his many thoughtful comments that substantially improved the presentation of these results.  Finally, we wish to thank the staff at NASA Pleiades, NASA Discover, and the Argonne Leadership Computing Facility (ALCF; particularly Wei Jiang), without whom these simulations would not have been possible.  

This work was supported by NASA grants NNX09AB04G, NNX14AC05G, NNX11AJ36G, and NNX14AG05G.  Featherstone and the development of $Rayleigh$ were further supported by the Computational Infrastructure for Geodynamics (CIG), which is supported by the National Science Foundation award NSF-094946.  This work used computational resources provided through NASA HEC support on the Pleiades and Discover supercomputers.  Further support was provided by an award of computer time on the Mira supercomputer through the Innovative and Novel Computational Impact on Theory and Experiment (INCITE) program.  This research used resources of the ALCF, which is a DOE Office of Science User Facility supported under Contract DE-AC02-06CH11357.

\input Appendices.tex

\input Bibliography.tex
\clearpage

\end{document}

%% file: definitions.tex
 {\begin{list}{}%
         {\setlength{\leftmargin}{#1}  
          \setlength{\rightmargin}{#1}}%
         \item[]%
 }
 {\end{list}}

\newcommand{\del}{\mbox{\boldmath $\nabla$}}

\newcommand{\cross}{\mbox{\boldmath $\times$}}

\def\avg{\bar}

\def\del{\nabla}
\def\cross{\times}
\def\avg{\bar}
\def\vec{\boldsymbol}
\def\scrD{\mathcal{D}}
\def\scrR{\mathcal{R}}

%% file: Tables.tex
\def\polytable{
\begin{table}[t]
\centering\small
\begin{tabular}[t]{ll}\\
\multicolumn{2}{c}{TABLE 1}\\
\multicolumn{2}{c}{Variable Input Parameters}\\\hline\hline
 Param & Value\\\hline
$M_i$    & 1.989$\times\mathrm{10}^{33}$ g\\
$\rho_i$ & 1.805$\times\mathrm{10}^{-1}$ g cm$^{-3}$\\
$c_p$    & 3.5$\times\mathrm{10}^8$ \\
$n$      & 1.5 \\
$N_\rho$ & 1,2,3,4 \\
$r_i$    & 5.0$\times\mathrm{10}^{10}$ cm \\
$r_o$    & 6.586$\times\mathrm{10}^{10}$ cm \\\hline
\end{tabular}
\tablecomments{\label{inputs_table} Polytropic reference state parameters for all cases.  The reference state prescription differs between simulations in the value of $N_\rho$ chosen only.}
\end{table}
}
\def\joneshydro{
\begin{table}[t]
\label{hydro_table}\centering\small
\begin{tabular}[t]{lrrr}\\
\multicolumn{4}{c}{TABLE A.1}\\
\multicolumn{4}{c}{Anelastic MHD Benchmark Results}\\\hline\hline
Observable & Rayleigh & Glatzmaier & \% Difference \\\hline

Kinetic Energy (erg)  &  5.56967$\times 10^{35}$ & 5.57028$\times 10^{35}$ & -1.09510$\times 10^{-2}$ \\
Zonal Kinetic Energy (erg)        &  6.37973$\times 10^{34}$ & 6.38099$\times 10^{34}$ & -1.97462$\times 10^{-2}$ \\ 
Meridional Kinetic Energy (erg)   &  1.49799$\times 10^{32}$ & 1.49825$\times 10^{32}$ & -1.73536$\times 10^{-2}$ \\ 
Entropy (erg g$^{-1}$ K$^{-1}$)        &  7.9449$\times 10^{5}$ & 7.9452$\times 10^{5}$ & -3.7759$\times 10^{-3}$ \\ 
$u_\phi$ (cm s$^{-1}$)            &  6.8638$\times 10^{2}$ & 6.9027$\times 10^{2}$ & -5.6355$\times 10^{-1}$ \\ 
Drift Frequency (rad s$^{-1}$) &  3.10509$\times 10^{-6}$ & 3.10512$\times 10^{-6}$ & -9.6615$\times 10^{-4}$ \\\hline

\end{tabular}
\tablecomments{ Hydrodynamic benchmark results for $Rayleigh$.  The simulation was run with a resolution of $128\times192\times384$ ($N_r\times N_\theta \times N_\phi$), corresponding to 96 Chebyshev modes and a maximum spherical harmonic degree of 127.  All quantities reported were averaged over 10,000 time steps following equilibration of the steady-state solution.  The time step size was 30 seconds, and the total evolution time was 4.8$\times10^6$ seconds.}
\end{table}
}

\def\jonesmhd{
\begin{table}[t]
\label{mhd_table}\centering\small
\begin{tabular}[t]{lrrr}\\
\multicolumn{4}{c}{TABLE A.2}\\
\multicolumn{4}{c}{Anelastic MHD Benchmark Results}\\\hline\hline
Observable & Rayleigh & Glatzmaier & \% Difference \\\hline
Kinetic Energy (erg) &  8.03633$\times 10^{36}$ & 8.03623$\times 10^{36}$ & 1.24436$\times 10^{-3}$ \\ 
Zonal KE (erg)       &  1.15320$\times 10^{36}$ & 1.15318$\times 10^{36}$ & 1.73433$\times 10^{-3}$ \\
Meridional KE (erg)  &  1.01585$\times 10^{33}$ & 1.01587$\times 10^{33}$ & -1.96876$\times 10^{-3}$  \\
Magnetic Energy (erg)&  6.13362$\times 10^{36}$ & 6.13333$\times 10^{36}$ & 4.72826$\times 10^{-3}$  \\
Zonal ME (erg)       &  4.62063$\times 10^{36}$ & 4.62046$\times 10^{36}$ & 3.67929$\times 10^{-3}$  \\
Meridional ME (erg)  &  3.24922$\times 10^{36}$ & 3.24927$\times 10^{36}$ &-1.53881$\times 10^{-3}$  \\
Entropy (erg g$^{-1}$ K$^{-1}$)        &  6.0886$\times 10^{5}$ & 6.0893$\times 10^{5}$ &-4.4340$\times 10^{-2}$  \\
$u_\phi$ (cm s$^{-1}$)           & -2.9374$\times 10^{3}$ & -2.9422$\times 10^{3}$ &-1.6314$\times 10^{-1}$ \\
$B_\theta$ (G)         &  2.7187$\times 10^{2}$ & 2.7292$\times 10^{2}$ &-3.8473$\times 10^{-1}$  \\
Drift Frequency (rad s$^{-1}$) &  4.3075$\times 10^{-6}$ & 4.3076$\times 10^{-6}$ &-2.3215$\times 10^{-3}$  \\\hline

\end{tabular}
\tablecomments{ Hydrodynamic benchmark results for $Rayleigh$.  The simulation was run with a resolution of $128\times192\times384$ ($N_r\times N_\theta \times N_\phi$), corresponding to 96 Chebyshev modes and a maximum spherical harmonic degree of 127.  All quantities reported were averaged over 10,000 time steps following equilibration of the steady-state solution.  The time step size was 200 seconds, and the total evolution time was 5$\times10^8$ seconds.}
\end{table}
}


%% file: Figures.tex
\def\refheat{
\begin{figure*}[t!]
\centerline{\epsfig{file=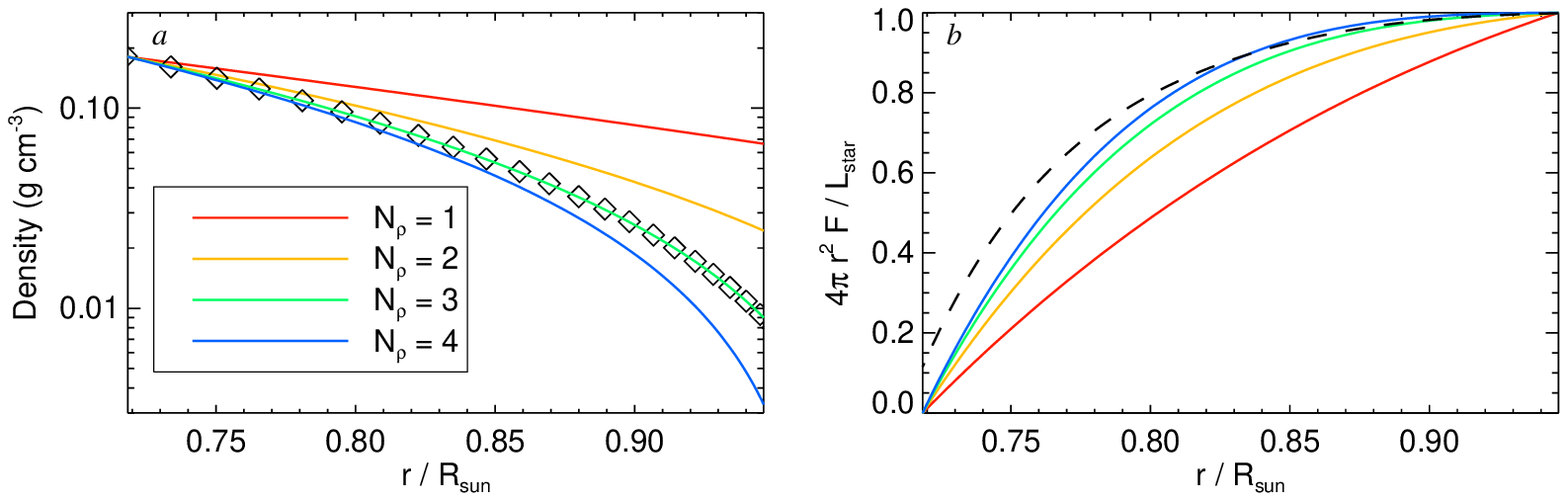,width=\textwidth}}
\caption{\footnotesize\label{fig:ref_heating}Polytropic reference states used in this study.  (\textit{a}) Radial variation of density corresponding to each value of $N_\rho$ used in our simulations (indicated by line coloring). The $N_\rho=3$ cases closely resemble the Model-S density stratification (black diamonds; Christensen-Dalsgaard et al. 1996).  (\textit{b})  Radial variation of thermal energy flux that must be transported by convection and thermal conduction for each value of $N_\rho$ (coloring as in panel \textit{a}).  Fluxes have been normalized by the stellar flux.  The flux corresponding to Model S is indicated by the dashed black line.  }
\end{figure*}
}

\def\kescaling{
\begin{figure*}[t!]
\vspace{-.1in}
\centerline{\epsfig{file=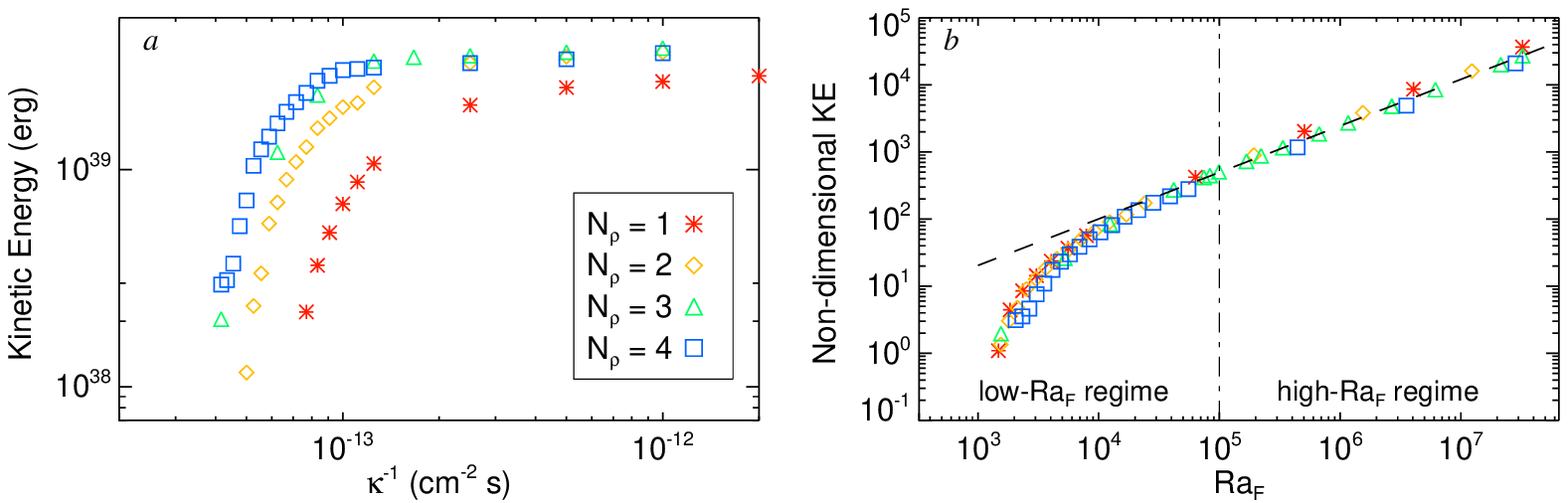,width=\textwidth}}
\vspace{-.1in}
\caption{\footnotesize\label{fig:ke_scaling}Dimensional and non-dimensional view of kinetic energy scaling. (\textit{a}) Dimensional kinetic energy $KE$ vs. $\kappa^{-1}$ for all cases run with $L_\star=L_\odot$.  Colored symbols indicate the level of density stratification.  At sufficiently low $\kappa$, $KE$ approaches a constant value within each $N_\rho$ series.  (\textit{b})  Non-dimensional kinetic energy $\widehat{KE}$ vs. flux Rayleigh number $Ra_F$ for \textit{all} cases (regardless of the value of $L_\star$).  Colored symbols indicate $N_\rho$ as in panel \textit{a}.  When viewed non-dimensionally, all $N_\rho$ series collapse onto a single curve and approach an asymptotic scaling law with $\widehat{KE}\propto Ra_F^{0.694}$ (dashed line).  The dilineation between high- and low-$Ra_F$ regimes is indicated by the dash-dotted line.}
\end{figure*}
}

\def\kespectra{
\begin{figure*}[p!]
\centerline{\epsfig{file=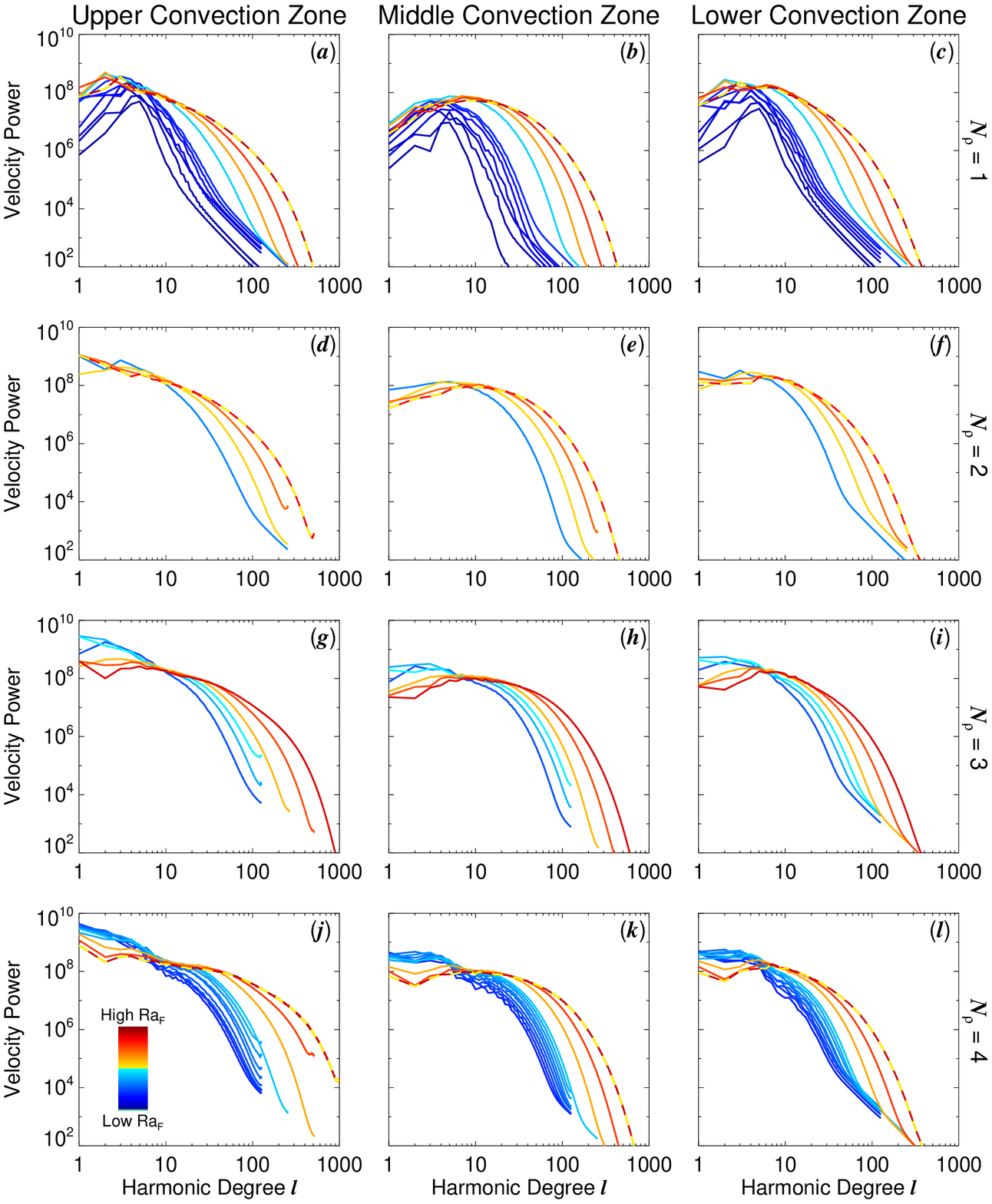,width=0.9\textwidth}}
\caption{\footnotesize\label{fig:ke_spectra}Time-averaged velocity power spectra (dimensional) for all cases run using $L_\star=L_\odot$.  Each row corresponds to a single value of $N_\rho$ (indicated on the right), and each column corresponds to a single depth within the convective shell.  Within each panel, spectra for all cases at that depth and $N_\rho$ are displayed.  Low-$Ra_F$ cases are indicated in blue tones, and high-$Ra_F$ cases in red tones.  The highest-$Ra_F$ case within each series is further indicated by the dashed, dark red line.  As $Ra_F$ is increased, power at low $\ell$-values increases initially.  At high $Ra_F$, it decreases as high-wavenumber power is generated at the expense of low-wavenumber power.  }
\end{figure*}
}

\def\fluxwidth{
\begin{figure*}[t!]
\centerline{\epsfig{file=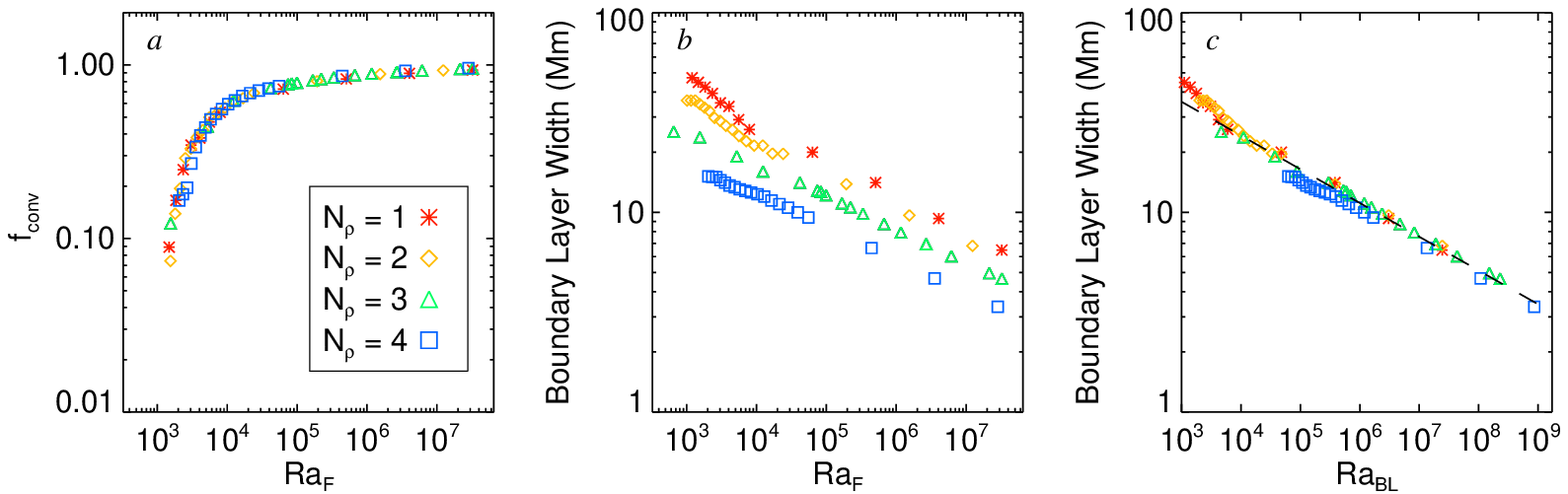,width=\textwidth}}
\caption{\footnotesize\label{fig:flux_width}Thermal boundary layer and energy transport scaling with Rayleigh number. (\textit{a}) Fractional convective flux $f_{conv}$ vs. $Ra_F$.  $f_{conv}$ approaches unity as $Ra_F$ is increased, indicating that conduction is playing a minimal role in the energy transport. (\textit{b}) Boundary layer width plotted vs. $Ra_F$ and (\textit{c}) the boundary-layer dependent Rayleigh number $Ra_{BL}$.  The boundary layer width follows a clear scaling law of $w_{BL}\propto Ra_{BL}^{-1/6}$ (dashed line, panel \textit{c}). }
\end{figure*}
}

\def\keradius{
\begin{figure*}[t!]
\centerline{\epsfig{file=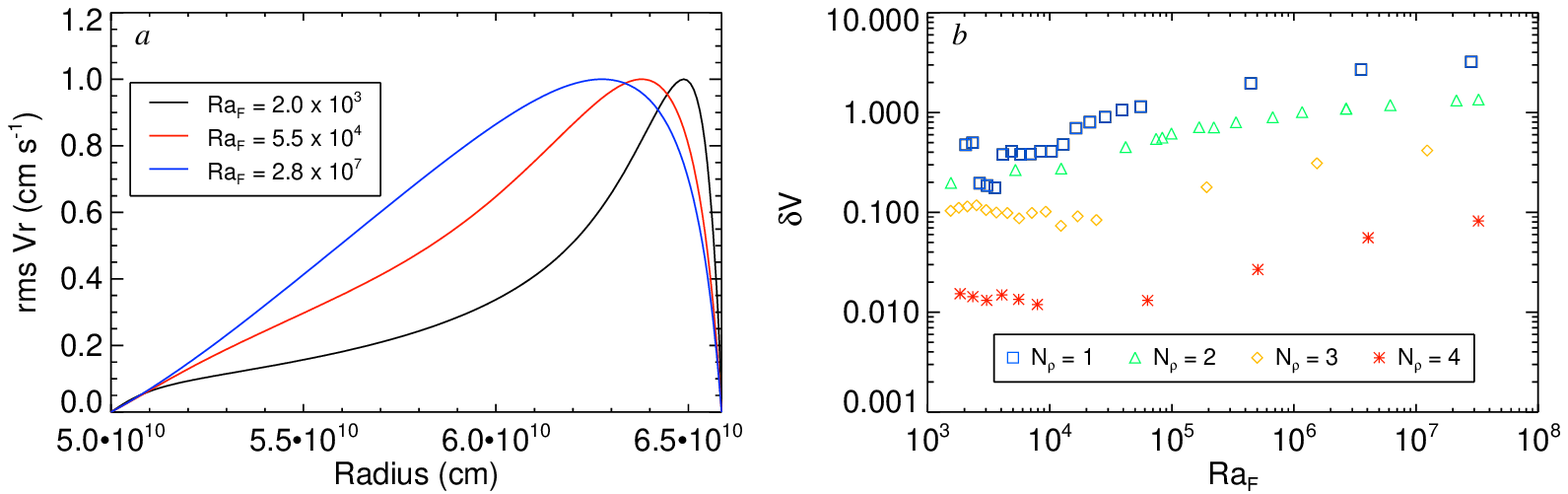,width=\textwidth}}
\caption{\footnotesize\label{fig:ke_radius}Near-surface and deep-convection-zone flow speed difference as a function of Rayleigh number and $N_\rho$. (\textit{a})  Normalized \textit{rms} radial velocity $V_r$ as a function of depth for $N_\rho=4$ and three values of $Ra_F$ (indicated by colored lines).  The difference in convective flow speed between mid and upper convection zone becomes more pronounced with increasing $Ra_F$. (\textit{b})  Relative speed difference $\delta V$ for all cases plotted as a function of Rayleigh number and $N_\rho$ (indicated by the colored symbols).  $\delta V$ increases with both $Ra_F$ and $N_\rho$ but is most sensitive to $N_\rho$.  }
\vspace{-0.2in}
\end{figure*}
}

%% file: Abstract.tex
\begin{abstract}
Convection plays a central role in the dynamics of any stellar interior, and yet its operation remains largely-hidden from direct observation.  As a result, much of our understanding concerning stellar convection necessarily derives from theoretical and computational models.  The Sun is, however, exceptional in that regard.  The wealth of observational data afforded by its proximity provides a unique testbed for comparing convection models against observations.  When such comparisons are carried out, surprising inconsistencies between those models and observations become apparent. Both photospheric and helioseismic measurements suggest that convection simulations may overestimate convective flow speeds on large spatial scales. Moreover, many solar convection simulations have difficulty reproducing the observed solar differential rotation due to this apparent overestimation.   We present a series of 3-dimensional (3-D) stellar convection simulations designed to examine how the amplitude and spectral distribution of convective flows are established within a star's interior.  While these simulations are non-magnetic and non-rotating in nature, they demonstrate two robust phenomena. When run with sufficiently high Rayleigh number, the integrated kinetic energy of the convection becomes effectively independent of thermal diffusion, but the spectral distribution of that kinetic energy remains sensitive to both of these quantities. A simulation that has converged to a diffusion-independent value of kinetic energy will divide that energy between spatial scales such that low-wavenumber power is overestimated, and high-wavenumber power is underestimated relative to a comparable system possessing higher Rayleigh number.  We discuss the implications of these results in light of the current inconsistencies between models and observations.
\end{abstract}

%% file: Introduction.tex
\section{Introduction}\label{sec:bfp}
Convection is a pervasive phenomenon within stars and serves as the principle transport mechanism by which fusion-generated energy is transmitted across significant portions of the stellar interior.   This transport may occur primarily within a convective core, as in the case of massive stars, or within an outer convective envelope, as in cool stars such as the Sun.  As that convection takes place within a rotating, electrically conductive plasma, it undoubtedly plays an active role in the generation of stellar magnetism. More specifically, rotation imbues convective motions with helicity which may serve to amplify magnetic fields through the so-called $\alpha$-effect (e.g., Moffat 1978). Further, the propensity of convective flows to redistribute angular momentum---thereby driving a differential rotation---may also render convective motion central to the generation of coherent, global-scale magnetic fields through the $\Omega$-effect (i.e., large-scale rotational shearing of the field).

Convection in the stellar context is thought to be highly turbulent; typical estimates for the Reynolds and Rayleigh numbers are 10$^{13}$ and 10$^{20}$ respectively (e.g., Ossendrijver 2003).  These extreme parameters make stellar convection challenging to study, whether experimentally or theoretically. This challenge is compounded by the fact that such convection occurs within a rotating, magnetized environment.

Investigations into stellar convection are most readily initiated within the context of the Sun.  The wealth of observational data afforded by the proximity of that star make it an ideal test bed against which predictions of stellar convection models may be tested and calibrated before extending those models to other stars.  Despite the abundance of solar data, the convective motions central to its dynamo remain largely hidden from view in all but the shallowest layers of the convection zone.  Whatever its detailed structure, that convection must satisfy two robust observational constraints: it must transport one solar luminosity's worth of energy across the convection zone, and it must efficiently redistribute angular momentum in order to sustain the Sun's helioseismically inferred differential rotation.

Exploring the magnetohydrodynamical balances at work in stellar interiors has largely been the purview of three-dimensional (3-D),  nonlinear, numerical convection simulations.  In the case of the Sun, these simulations are often run by assigning solar values to most model variables, notably the rotation rate and the luminosity, while admitting diffusivities that are computationally tractable but also necessarily orders of magnitude larger than their intended microscopic counterparts (e.g., Brun et al. 2004; Ghizaru et al. 2010; Hotta et al. 2014). It is now becoming increasingly evident that such quasi-laminar simulations have succeeded in maintaining a solar-like differential rotation and energy flux in part due to the out-sized role played by diffusion in many of those models.

As advances in computational power have enabled the use of lower and ostensibly more stellar-like diffusivities, inconsistencies between models and observations have begun to arise.  Notable among these are the so-called anti-solar states of differential rotation which exhibit an equator of retrograde rotation and polar regions of prograde rotation. Such states develop when the characteristic convective turnover time drops below some critical fraction of the rotation period (e.g., Gilman 1977; Glatzmaier \& Gilman 1982).  This transition occurs naturally in solar simulations as the level of diffusion is decreased (Featherstone \& Miesch 2015), and it tends to arise even in those models that do not employ an explicit, Fickian diffusivity  (Hotta et al. 2014).  While almost certainly relevant within a broader stellar context (see e.g., K\H{o}v\'{a}ri et al. 2007), these anti-solar states stand at odds with the well-established internal rotation profile of the Sun (e.g., Thompson et al. 2003). 

Further inconsistencies arise when the distribution of velocity power among different spatial scales is examined in detail. Photospheric observations and helioseismic observations of the deeper interior both suggest that numerical simulations may over-estimate the amplitude of the solar convection on large spatial scales (Hanasoge et al. 2012; Lord et al. 2014).  The helioseismic consensus concerning deep flows remains murky, however.  Recent ring-analysis measurements of subsurface flows throughout the near-surface sheer layer exhibit good agreement with convection models (Greer et al. 2015).  While the question of subsurface flow speeds remains an open question, we note that the time-distance measurements of Hanasoge et al. (2012), if confirmed, are particularly troubling as they suggest that convection on spatial scales larger than 70 Mm can be at most 5--6 m s$^{-1}$.  This stands in contrast to theoretical estimates based on gyroscopic pumping which suggest that convective flows must be at least 30 m s$^{-1}$ on those spatial scales (Miesch et al. 2012).  Weaker flows would have difficulty generating Reynolds stresses which are strong enough to maintain the solar differential rotation.  

These theoretical and observational disagreements suggest that something fundamental, related to the transport of heat and angular momentum by magnetized convection, may be missing in stellar convection models.  Until these issues can be resolved in the context of the Sun, for which we have such an abundance of observational data, the relationship of {\sl any} stellar convection model to an actual star remains highly questionable.
 
We present a set of numerical experiments designed to examine how the structure and amplitude of the convective flow field is determined within a stellar interior.  Specifically, we ask, \textit{``How is the amplitude and spectral-power distribution of stratified convection related to the stellar luminosity and level of diffusion present in the system."} For this paper, we address that question without considering the effects of rotation and magnetism which, while admittedly important, would substantially extend the parameter space of this already challenging study.  We choose instead to focus our initial study on how density stratification, luminosity, and diffusivity relate to the amplitude and spatial power spectrum of the resulting convection. Our results demonstrate that the convective kinetic energy is related to these three parameters through an unambiguous scaling relationship.  In addition, we demonstrate that models run in the typical stellar-convection fashion may naturally overestimate low-wavenumber power in the convective power spectrum.

This paper is organized as follows. In \S\ref{sec:model} and \S\ref{sec:experiment}, we discuss our numerical model and the parameter space explored in our numerical experiments.  We then discuss our results in \S\ref{sec:results}, followed by a discussion of their implications in \S\ref{sec:conclusions}.

%% file: Model.tex
\section{Numerical Model}\label{sec:model}
This study is based around a series of 3-D, nonlinear convection models run using the $Rayleigh$ convection code, developed as a community code for the Computational Infrastructure for Geodynamics (CIG).  $Rayleigh$ simulates convection using the spectral transform approach described in Glatzmaier (1984).  We employ a spherical geometry and represent the horizontal variation of all variables along spherical surfaces using spherical harmonics $Y_\ell^m(\theta,\phi)$.  Here, $
\ell$ is the spherical harmonic degree, and $m$ is the azimuthal mode order.  In the radial direction, we employ a Chebyshev collocation method, expanding all variables using  Chebyshev polynomials $T_n(r)$, where $n$ is the degree of the Chebyshev polynomial.  Derivatives are calculated accurately within the spectral representation by exploiting the properties of these two sets of basis functions.  We work on a de-aliased grid, such that the number of collocation points in each dimension is larger than the corresponding number of spectral modes by a factor of 3/2.

As our approach is pseudo-spectral in nature,  nonlinear terms are calculated in physical space after first transforming the relevant variables from the spectral configuration.  Time-integration is carried out in the spectral configuration using a hybrid implicit-explicit approach.  A Crank-Nicolson method is used for linear terms, and an Adams-Bashforth approach is employed for the nonlinear terms.  Both components of the time-stepping possess 2nd-order accuracy.  Algorithmically, it is the parallelization of $Rayleigh$, which parallelizes efficiently on up to O(10$^5$) cores,  that differs significantly from that of Glatzmaier (1984).  We have verified our particular implementation of that approach against two established benchmark tests, and we provide results from those tests in Appendix A. 

Our study is concerned with convection as it manifests deep within stellar interiors, far removed from the photospheric surface where radiative processes may contribute considerably to the energetics of the convection.  In such a region of the star, plasma motions are subsonic and perturbations to thermodynamic variables are small compared to their mean, horizontally-averaged values.  Under such conditions, the anelastic approximation, which we choose to employ, provides a convenient means of describing the system's thermodynamics (Gough 1969; Gilman \& Glatzmaier 1981).

When using this approximation, thermodynamic variables are linearized about a spherically symmetric, time-independent reference state with density $\avg{\rho}$, pressure $\avg{P}$, temperature $\avg{T}$, and specific entropy $\avg{S}$.  Fluctuations about this state are denoted, without overbars ($\rho$, $P$, $T$, and $S$).  A further consequence of the anelastic approximation is that the mass flux is solenoidal, reducing the continuity equation to
\begin{equation}  
  \label{eq:continuity}
  \del \cdot(\avg{\rho}\vec{v}) = 0,
\end{equation}
where $\vec{v} = (v_r,v_{\theta},v_{\phi})$ is the velocity vector expressed in spherical coordinates.  The lack of any time derivative in Equation (\ref{eq:continuity}) means that sounds waves are naturally filtered out as a consequence of this approach.  The divergence-free constraint for the mass flux is enforced by projecting $\vec{v}$ onto poloidal and toroidal streamfunctions ($W$ and $Z$ respectively), such that
\begin{equation}  
	\avg{\rho}\vec{v} = \del\cross\del\cross(W\vec{e}_r)+\del\cross(Z\vec{e}_r).
\end{equation}
The unit vector in the radial direction is indicated by $\vec{e}_r$.  The momentum equation is given by
\begin{equation}  
  \label{eq:momentum}
\begin{split}
  \avg{\rho}\frac{D\vec{v}}{Dt}
  =
  -\avg{\rho}\del\frac{P}{\avg{\rho}} - \frac{\avg{\rho}S}{c_p} \vec{g} - \del \cdot \scrD,
\end{split}
\end{equation}
where $\vec{g}$ is the gravitational acceleration, and $c_p$ is the specific heat at constant pressure.  In writing the momentum equation this way, we have employed the so-called Lantz-Braginsky-Roberts approximation, which is exact for adiabatic references states such as those employed in this study (Lantz 1992; Braginsky \& Roberts 1995).  The viscous stress tensor $\scrD$ is given by 
\begin{equation}  
\label{eq:stress}
  \scrD_{ij} = -2 \avg{\rho} \nu \left[e_{ij}
    - \frac{1}{3}(\del \cdot \vec{v})\delta_{ij} \right],
\end{equation}
where $e{_{ij}}$ is the strain rate tensor. The kinematic viscosity is denoted by $\nu$, and $\delta_{ij}$ is the Kronecker delta. Our thermal energy equation is given by
\begin{equation} 
  \label{eq:entropytwo}
  \begin{split}
  \avg{\rho}\avg{T}\frac{DS}{Dt} = 
  \del \cdot [\kappa \avg{\rho} \avg{T} \del S] 
                    + 2 \avg{\rho}\nu \left[e_{ij}e_{ij} - \frac{1}{3}(\del \cdot
\vec{v})^2\right]+Q,
  \end{split}
\end{equation}
where the thermal diffusivity is denoted by $\kappa$.  Sources of internal heating and cooling are encapsulated by the functional form of $Q$.  A linearized equation of state closes our set of equations and is given by
\begin{equation} 
  \frac{\rho}{\avg{\rho}} = \frac{P}{\avg{P}} - \frac{T}{\avg{T}}
    =  \frac{P}{\gamma \avg{P}} - \frac{S}{c_p},
\end{equation}
assuming the ideal gas law
\begin{equation} 
  \avg{P} = \scrR \avg{\rho} \avg{T}.
\end{equation}
The specific heat at constant pressure is denoted by $c_p$, $\scrR$ is the gas constant, and $\gamma$ is the adiabatic index of the gas.  

\section{The Numerical Experiment}\label{sec:experiment}
We have constructed a series of 63 model stellar convection zones designed to explore how the convective kinetic energy depends on three model parameters: the degree of density stratification, level of thermal diffusion, and the stellar luminosity.  A detailed listing of all model parameters may be found in Appendix B. 

Each of our models is constructed using a polytropic background state following Jones et al. (2011).  This approach has the advantage that the thermodynamic background may be specified analytically, making it easily reproducible.  Our background may be completely specified by seven numbers: the inner radius of the shell $r_i$, the outer radius of the shell $r_o$, the polytropic index $n$, the number of density scale heights occurring within the shell $N_\rho$, the mass interior to the shell $M_i$, the density at the inner boundary $\rho_i$, and the specific heat $c_p$.  With the exception of $N_\rho$, these values are the same for all experiments reported on here, and we list them in Table 1.  All background states are adiabatically stratified, possessing a polytropic index $n$ of 1.5 and an adiabatic index $\gamma$ of 5/3. Further details of the polytropic background are provided in Appendix C for completeness.  

\polytable

We have chosen to examine a range of density stratifications ($N_\rho$) for this study.  One consequence of this approach is that shells with a higher degree of density stratification possess a lower mass than their more weakly stratified counterparts.  We note that when constructed using the parameters from Table 1, the thermodynamic background state closely resembles that of the Sun for the value $N_\rho=3$.  We plot the density variation with radius for each of our four chosen values of $N_\rho$, along with that from a standard solar model (Model S; Christensen-Dalsgaard et al. 1996), in Figure \ref{fig:ref_heating}\textit{a}.

\refheat

For all simulations, we have adopted impenetrable and stress-free boundary conditions such that
\begin{equation}
\label{eq:vboundary}
v_r(r = r_i,r_o)=\left.\frac{\partial (v_{\theta}/r)}{\partial r}\right|_{r = r_i,r_o}=\left.\frac{\partial (v_{\phi}/r)}{\partial r}\right|_{r = r_i,r_o}=0.
\end{equation}
The radial entropy gradient is forced to vanish at the lower boundary of the convection zone, and the entropy perturbations are forced to vanish at the upper boundary, with
\begin{equation}
\label{eq:sboundary}
\left.\frac{\partial S}{\partial r} \right|_{r=r_i}= 0,~~S(r_o) = 0.
\end{equation}
Thus, there is no diffusive entropy flux across the lower boundary.  Instead, heat enters the system through internal deposition by $Q$, which drops to zero at the upper boundary.  In all simulations, we adopt a functional form of $Q$ that depends only on the background pressure profile such that
\begin{equation} 
\label{eq:heating}
Q(r,\theta,\phi) = \alpha (\avg{P}(r)-\avg{P}(r_{o})).
\end{equation}     
The normalization constant $\alpha$ is chosen so that
\begin{equation}
\label{eq:heating}
L_\star = 4\pi \int_{ri}^{ro} Q(r) r^2 dr, 
\end{equation}
where $L_\star$ is the stellar luminosity.  The thermal energy flux $F(r)$ that convection and conduction must transport across a spherical surface at radius $r$ is then given by
\begin{equation}
\label{eq:netflux}
F(r) = \frac{1}{r^2}\int_{ri}^{r} Q(x) x^2 dx.
\end{equation}
The value of $L_\star$ is varied between simulations, as is the value of $N_\rho$, and that of the diffusion coefficients $\nu$ and $\kappa$.  In all instances, we have adopted a Prandlt number $Pr=\nu/\kappa$ of unity. 

The functional form of $Q$ means that as the value of $N_\rho$ is increased, the heating becomes more focused near the lower boundary.  This behavior is illustrated in Figure \ref{fig:ref_heating}\textit{b}, where we have plotted $F(r)$, normalized by the stellar flux at each radius.  Our prescription for the internal heating in the $N_\rho=3$ cases resembles that derived using solar temperature and opacity values tabulated in Model S (dashed line). We note that small differences, comparable to those observed between reference states with differing values of $N_\rho$, do exist between the $N_\rho=3$ case and the solar model.  Those differences arise primarily from the fact that we have chosen the base of our model convection zones to lie at a radius of 5$\times10^{10}$ cm, somewhat higher than that of the Sun.  The results discussed in \S\ref{sec:results} suggest, however, that the functional form of the internal heating has little effect on the kinetic energy realized in our high-Rayleigh-number simulations.  

At the upper boundary, heat exits the system via thermal conduction.  There, the steepness of the equilibrated entropy gradient is entirely dependent upon the thermodynamic background state, the value of the thermal diffusivity $\kappa$, and the chosen luminosity. Specifically, the time-averaged value of the entropy gradient must equilibrate to
\begin{equation}
\label{eq:sgradient}
\left.\frac{\partial S}{\partial r}\right|_{r=r_o} = \frac{L_\star}{4\pi r_o^2\kappa\rho(r_o)T(r_o)}
\end{equation}

\subsection{Non-dimensionalization}
All simulations were run in a manner analogous to one commonly used in solar convection studies (e.g., Brun et al. 2004; Ghizaru et al. 2010; Hotta et al. 2014, Featherstone \& Miesch 2015).  Namely, we have run these simulations using \textit{dimensional} parameters for the equation coefficients, and we assign ``stellar'' values to those parameters, with the notable exception of the diffusivities.  The values chosen for our diffusion coefficients are instead motivated purely by matters of computational feasibility; thus they are necessarily orders of magnitude larger than the microscopic values.  It is  customary to suggest that this prescription for the diffusivity constitutes a form of subgrid-scale model, representing the mixing of heat and momentum by small-scale eddy motions that cannot be sufficiently resolved on the coarse computational grid.  We make no such assertion here, and remark simply that our diffusivities are rather large when compared to stellar values.  The diffusion coefficients $\nu$ and $\kappa$ are taken to be constant values within each of our simulations.  In particular, they possess no variation in radius.

Across our set of numerical convection experiments, we have varied the degree of thermal stratification (characterized by $N_\rho$), the luminosity of the star $L_\star$, and the thermal diffusivity $\kappa$.  This three-dimensional parameter space may be collapsed by considering an appropriate non-dimensionalization of the system.  As we have chosen a $Pr$ of unity and have neglected the effects of rotation and magnetism, all runs may be well-characterized by a single non-dimensional number, the Rayleigh number.

For Boussinesq convection, the Rayleigh number is typically defined in terms of the temperature, but in our anelastic formulation, the thermal properties of the fluid are instead conveyed by entropy.  Our simulations also differ from the classical Rayleigh-B\'{e}nard setup, which possesses fixed-temperature boundaries, in that we have effectively imposed a flux by specifying $Q$ and the entropy boundary conditions of Equation (\ref{eq:sboundary}).  It is thus appropriate for our system to define a flux Rayleigh number in terms of entropy.  Throughout this paper, the term \textit{Rayleigh number} will be used to refer to a \textit{flux} Rayleigh number $Ra_F$, which we define as
\begin{equation}
\label{eq:ranum}
\mathrm{Ra}_F = \frac{\widetilde{g}\widetilde{F}H^4}{c_p\widetilde{\rho}\widetilde{T}\nu\kappa^2}.
\end{equation}
We choose to use volume averages for all quantities possessing a tilde in the definition of $Ra_F$, making it a bulk Rayleigh number.  Here \textit{H} is a typical length scale, and we choose the shell depth $r_o-r_i$ as its value.  Our numerical experiments span the range $10^3 \le Ra_F \le 3\times10^7$.  Each simulation was initialized using a small random thermal perturbation, evolved until the kinetic energy reached a statistically steady state, and further evolved for at least one thermal diffusion time further following the onset of that statistically steady state.

%% file: Results.tex
\section{Survey of Results}\label{sec:results}

\subsection{Kinetic-Energy Scaling and Spectral Distribution}
\kescaling
We begin our discussion of the convective energetics by looking at the integrated kinetic energy $KE$, defined as
\begin{equation}
\label{ke_def}
KE = \frac{1}{2}\int_{V}\avg{\rho}(r)\left|\vec{v}(r,\theta,\phi)\right|^2dV.
\end{equation}
The relationship of $KE$ to thermal diffusivity is illustrated in Figure \ref{fig:ke_scaling}$a$.  There, we have plotted $KE$ vs. $\kappa^{-1}$, using colored symbols to distinguish between the level of stratification employed in each model.  Each data point represents an average of 3000 days of simulation time.  This is approximately one thermal diffusion time across the shell for those cases with $\kappa \ge 1\times10^{12}$ cm$^2$ s$^{-1}$.  Only two of our models possess a lower value of $\kappa$ ($5\times10^{11}$ cm$^2$ s$^{-1}$), but this averaging interval still encompasses several tens of convective overturning times.   We have also omitted those cases run with a non-solar value of $L_\star$ in this figure so that the general trend may be clearly seen.  For each set of curves, the integrated $KE$ rises until some critical value of $\kappa$ is reached.  Below this critical value of $\kappa$, $KE$ asymptotes to a nearly constant value that is \textit{essentially independent of the level of diffusion $\kappa$}.  These simulations have thus reached a state of so-called turbulent free-fall (e.g., Spiegel 1971), which we discuss further in \S\ref{sec:conclusions} of this paper.  A fit to the low-$\kappa$ data in Figure \ref{fig:ke_scaling}$a$ yields the relationship
\begin{equation}
\label{eq:kappa_dep}
KE \propto \kappa^{-0.082 \pm 0.024},
\end{equation}
which is only weakly dependent on $\kappa$.

The point of transition to this asymptotic regime depends on $N_\rho$.  More weakly statified cases transition to the asymptotic regime at a lower value of $\kappa$ than their more strongly stratified counterparts.  In addition, the saturation level of $KE$ may be weakly dependent on $N_\rho$, as evinced by the $N_\rho = 1$ curve (red).  It is possible, however, that those weakly stratified cases are simply converging more slowly to the same value as the $N_\rho > 1$ cases.  All cases with $N_\rho>1$ converge to a similar value of $KE$.

The similarity of behavior in $KE$ between systems with different $N_\rho$ is made more readily apparent by examining how $KE$ depends on the Rayleigh number $Ra_F$.  In so doing, we choose a non-dimensional measure of the kinetic energy $\widehat{KE}$ such that
\begin{equation}
\label{eq:kend}
\widehat{KE} \equiv \frac{H^2}{\kappa^2 M} KE.
\end{equation}
The nondimensionalization has been carried out using the depth of the domain, the thermal diffusion timescale, and the mass $M$ contained within the spherical shell (which is a function of $N_\rho$).  In Figure \ref{fig:ke_scaling}$b$, we have plotted $\widehat{KE}$ vs. $Ra_F$.  When plotted in this fashion, all results fall along a similar curve that describes a power-law relationship beween $KE$ and $Ra_F$.  Performing a least-squares fit to the high-$Ra_F$ portion of this plot ($Ra_F \ge 10^5$), we find that
\begin{equation}
\label{eq:kescale}
\widehat{KE} \propto Ra_F^{0.694 \pm 0.008}.
\end{equation} 
The measured value of the exponent is very close to $\frac{2}{3}$.  This value indicates that beyond $Ra_F$ of 10$^5$, which we designate as the ``high-$Ra_F$'' region of this curve, diffusion is no longer playing a substantial a role in determining the convective amplitudes.  

While achieving a truly ``stellar'' value for the Rayleigh number is clearly impossible in a modern convection simulation, these results nevertheless imply a clear prerequisite for comparing any convection simulation against stellar observations.  \textbf{Namely, a simulation's parameters should place it well within the high-$\boldsymbol{Ra_F}$ region of Figure \ref{fig:ke_scaling}$b$ before any comparison is made against actual data.}

\kespectra

This result has interesting consequences for the relative spectral distribution of kinetic energy between high- and low-$Ra_F$ systems.  In Figure \ref{fig:ke_spectra}, we have plotted the kinetic-energy spectra for those cases in our study that possess the solar luminosity.  Spectra are plotted at three depths in each simulation (1 depth per column).  These depths are taken near the upper boundary, the mid-shell, and near the lower boundary. Different values of $N_\rho$ correspond to different rows.  Within each panel, all spectra for high-$Ra_F$ runs at that $N_\rho$ have been plotted.  Low values of $Ra_F$, indicated by dark blue tones, have been plotted for the $N_\rho=1$ cases in order to illustrate the trend for that region of parameter space.  They have been largely omitted for other values of $N_\rho$ to enhance legibility of the high-$Ra_F$ behavior.  The highest-Rayleigh-number spectrum in each panel is indicated by a dashed, dark red line.  Each spectrum represents a time-average over approximately 1,000 days of simulated time, corresponding to several tens of convective overturnings.

A systematic trend is visible across all cases plotted here.  As $Ra_F$ increases from an initially very low value, the integral of the kinetic-energy spectrum increases, as does the point-wise amplitude at roughly all wavenumbers.  This trend is commensurate with the low-$Ra_F$/high-$\kappa$ behavior depicted in Figures \ref{fig:ke_scaling}$a,b$.  At sufficiently high Rayleigh number, however, power in the high-$\ell$ portion of the spectrum continues to increase, but the low-wavenumber end begins to diminish in power.  The spectrum with the highest $Ra_F$ in each panel (red dashed line) thus has less power at large scales than its low-$Ra_F$ counterparts.  The ``low-wavenumber" end of the spectrum appears to occur in the approximate range $0 \leq \ell \leq 10$ for most simulations in this study.  

This phenomenon arises naturally from the fact that at sufficiently high Rayleigh number (and sufficiently low $\kappa$), the integrated kinetic energy becomes essentially independent of the degree of diffusion.  At the same time, as $\kappa$ is decreased and $Ra_F$ is increased, the flow field becomes more turbulent and develops structure on increasingly smaller scales.  Because the integral of the power spectrum is conserved, {\bf high-wavenumber power can only come at the expense of low-wavenumber power.}  This general trend is independent of depth and appears so be somewhat more pronounced for systems possessing larger values of $N_\rho$.

Before proceeding, we wish to emphasize that the results encapsulated by Figures \ref{fig:ke_scaling} and \ref{fig:ke_spectra} constitute the main points of this paper.  In summary:
\begin{enumerate}
\item When a global stellar convection simulation is not run with sufficiently high Rayleigh number, its resulting convective flow speeds will depend on the (unphysically high) value of $\kappa$, in addition to stellar parameters such as mass and luminosity.  A simulation that is not in the high-$Ra_F$ regime of Figure \ref{fig:ke_scaling}$b$ can be expected to yield $rms$ convective amplitudes which are systematically lower than those expected in an actual star.
\item For sufficiently turbulent (high-$Ra_F$) convection simulations, this problem does not arise.  The time-averaged, integrated kinetic energy approaches a constant value, independent of the level of diffusion employed.
\item Even in this asymptotic regime, models that correctly capture the integrated kinetic energy \textit{may naturally overestimate low-wavenumber power} because their Rayleigh number is still orders of magnitude smaller than that of an actual star.  Thus, simulated convective features that derive from large-scale motions may bear little relation to their intended stellar counterparts.
\end{enumerate}

\keradius

Changes in Rayleigh number also impact the spatial distribution of kinetic energy.  This is most clear when considering the variation of radial velocity with depth.  We have plotted the normalized $rms$ vertical velocity $V_r$ in Figure \ref{fig:ke_radius}$a$.  We define this quantity as 

\begin{equation}
\label{eq:vprime}
V_r(r)\equiv \frac{u_{rms}(r)}{\mathrm{max}\, u_{rms}},
\end{equation}
where $u_{rms}$ is the $rms$ vertical velocity at each radius.  The mean in the $rms$ is spatial and temporal; it has been taken over spherical shells and over 1,000 days of simulation time.  As the Rayleigh number is increased, the variation in radial velocity across the domain increases and tends to peak near the upper boundary.  We suggest that this behavior is the likely result of plumes being launched from the upper boundary layer whose thickness and gradient are both changing with Rayleigh number.  At high $Ra_F$, the boundary layer is characterized by a sharp entropy gradient, and is thus amenable to the generation of rapid, small-scale downflow plumes.   The location and width of the velocity peak are thus an indirect measure of the effective penetration depth of these plumes.  

This effect is most pronounced at high $Ra_F$ and high $N_\rho$, and we may quantify it by examing the relative velocity jump $\delta V$, which we define as
\begin{equation}
\label{eq:delv}
\delta V \equiv \frac{\mathrm{max}\,u_{rms} - u_{rms}(r_{mid})} {u_{rms}(r_{mid})},
\end{equation}
where $r_{mid}$ is the radial depth halfway between $r_i$ and $r_o$.  The relative velocity jump measured for each case is plotted in Figure \ref{fig:ke_radius}$b$.  The trend present in panel $a$ of that figure is more clearly illustrated here.  $\delta V$ increases with $Ra_F$ for all values of $N_\rho$ studied, reaching the highest values for the $N_\rho=4$ cases.  We note that it is due to the logarithmic axes that $\delta V$ appears to be asymptoting to a constant value within each $N_\rho$ series.  While the growth of $\delta V$ is slowing at high $Ra_F$, this quantity is still clearly increasing within each series.  We note further that we have not succeeded in identifying a clear scaling law linking the behavior of $\delta V$ to the value of $Ra_F$.  Nevertheless, the apparent trend suggests that a star such as the Sun, whose convection zone is characterized by much higher values of both $N_\rho$ and $Ra_F$, should possess a $\delta V$ greater than 4.  This behavior has implications for the solar near-surface shear layer, as we discuss in \S \ref{sec:conclusions}.

\subsection{Energy Transport}
Rotation and magnetism aside, the convective dynamics realized in our simulations will likely differ significantly from those found within a star due to both the large values of diffusivity we employ and the conductive boundary layer that develops near the top of the domain.  We have seen that diffusion plays a minimal role in the determination of the convective kinetic energy, and we now seek to quantify its role in the energy transport across the layer.  The thermo-mechanical transport of energy in our system may be characterized by three radial energy fluxes, namely the enthalpy flux $F_e$, the kinetic energy flux $F_{KE}$, and the conductive flux $F_c$, which we define as

\begin{equation}
\label{eq:fe}
F_e = \avg{\rho}c_p  \left<v_r T\right>,
\end{equation}

\begin{equation}
\label{eq:fke}
F_{KE} = \frac{1}{2}\avg{\rho} \left<v_r \left|\vec{v}\right|^2\right>,
\end{equation}
and
\begin{equation}
\label{eq:fc}
F_c = \kappa\avg{\rho}\avg{T}\left<\frac{\partial S}{\partial r}\right>
\end{equation}
respectively.  We consider averages of these quantities taken over 1,000 days of simulation time and over spherical surfaces, denoting that average by angular brackets.  Thus $F_e$, $F_{KE}$, and $F_c$ are a function of radius only.  At each radius, the sum of these three fluxes must equate to the net flux $F$, established by the internal heating $Q$ (Equation \ref{eq:netflux}), such that

\begin{equation}
\label{eq:ftotal}
F(r) = F_e(r) + F_{KE}(r)+F_c(r).
\end{equation}
We may now quantify the contribution of conduction by considering the fractional convective flux $f_{conv}$, which we define as
\begin{equation}
\label{eq:fconv}
f_{conv} \equiv \frac{\int_{V}F_e+F_{KE}\,dV}{\int_V F\,dV}= 1-\frac{\int_V F_{cond}\,dV}{\int_V F\,dV}.
\end{equation}
A value of zero for $f_{conv}$ indicates a lack of convective heat transport.  A value of $f_{conv}$ near unity indicates that convection plays a dominant role over thermal conduction.  The variation of time-averaged $f_{conv}$ with Rayleigh number is plotted for all simulations in Figure \ref{fig:flux_width}$a$.  This quantity is clearly asymptoting toward unity as $Ra_F$ is increased and reaches a value of 0.958 for our highest $Ra_F$, $N_\rho=4$ run.  Thermal conduction is thus playing a minimal role in the average transport of thermal energy at high Rayleigh number.   

\fluxwidth

The role of conduction may be further characterized by considering the extent of the thermal boundary layer.  We define the thermal boundary layer width in terms of the time-averaged mean entropy, such that
\begin{equation}
\label{eq:width}
w_{BL} = \frac{1}{\mathrm{max}\left<S(r)\right> }\int \left<S(r)\right>dr.
\end{equation}
The variation of $w_{BL}$ with Rayleigh number is illustrated in Figure \ref{fig:flux_width}$b$.  

The boundary layer width decreases systematically with increasing $Ra_F$ and exhibits a clear dependence on the degree of density stratification $N_\rho$.  At comparable $Ra_F$, simulations with low values of $N_\rho$ possess a wider thermal boundary layer than those with higher values of $N_\rho$.  This difference may be largely understood by noting that the thermal boundary layer is an upper-convection zone phenomenon and is better described by properties local to that region.  

We may define an alternative formulation of the Rayleigh number, $Ra_{BL}$, by substituting point-wise values taken at the upper boundary for the volume-averaged quantities in Equation (\ref{eq:ranum}).  Using this definition of the Rayleigh number, the multiple curves apparent in Figure \ref{fig:flux_width}$b$ nearly collapse onto a single curve, as depicted in Figure \ref{fig:flux_width}$c$.  A fit to the scaling exponent yields the (unsurprising) relationship
\begin{equation}
\label{eq:wscale}
w_{BL} \propto Ra_{BL}^{-0.169} 
\end{equation}
In other words, the boundary layer width scales very nearly as $\sqrt{\kappa}$.  We note that for our highest Rayleigh number run, the boundary layer width is roughly 3 Mm, or about 2\% of the shell depth.  Thus, while our thermal boundary layers lack the radiative processes at work in the Sun, their physical extent is confined to a similarly small region of the convective domain.


%% file: Conclusions.tex
\section{Perspectives and Conclusions}\label{sec:conclusions}
Our results hold interesting consequences for several aspects of convection zone dynamics.  Many of these results will undoubtedly be sensitive to the effects of rotation and magnetism which were omitted in this study.  Nevertheless, we discuss some implications of this work and remark that we are now pursuing a complementary set of simulations that incorporate those effects.
 
\subsection{Interpretation of the Kinetic Energy Scaling}
The kinetic-energy scaling observed in these simulations may be understood in terms of a free-fall argument (e.g., Spiegal 1971).  We outline such an argument as follows, assuming that $Pr=1$ (as is true for the models presented here) and that the kinetic energy derives from the potential energy $PE_{BL}$ associated with the boundary layer.  

Our conductive boundary layer possesses an entropy gradient given by Equation (\ref{eq:sgradient}) and a width that is roughly proportional to $Ra_{BL}^{-1/6}$ (i.e., $\sqrt{\kappa}$ ).  Thus, the entropy contrast $\Delta S$ across that layer should scale as
\begin{equation}
\label{eq:scaling1}
\Delta S \approx \left .\frac{\partial S}{\partial r}\right|_{r=r_o}  w_{BL} \propto \frac{L_\star \,Ra_{BL}^{-1/6}}{4\pi r_o^2\kappa\avg{\rho}(r_o)\avg{T}(r_o)}.
\end{equation}
The boundary layer possesses an entropy deficit with respect to the adiabatic interior of roughly $\Delta S/2$.  This entropy deficit translates into an overdensity within that region, with an associated mass $M_{BL}$ given by
\begin{equation}
\label{eq:scaling2}
M_{BL} \approx \frac{\avg{\rho}(r_o)}{c_p}\,\frac{\Delta S}{2} \times 4\pi r_o^2 \,w_{BL} \propto \frac{L_\star \, Ra_{BL}^{-1/3}}{2\kappa c_p\avg{T}(r_o)}
\end{equation}
The potential energy, and thus the kinetic energy associated with the convection then scales as
\begin{equation}
\label{eq:scaling3}
KE \approx PE_{BL} \approx M_{BL}\,\,g\,H \propto  \left(\frac{L_\star\, g }{c_p\,\avg{T}(r_o)}\right)^{2/3} \left(\frac{4\pi r_o^2 \, \avg{\rho}(r_o)}{H} \right)^{1/3}   ,
\end{equation}
which is independent of the thermal diffusivity owing to the two factors of $\sqrt{\kappa}$ arising from the boundary layer thickness.  As we observe only a very weak dependence on diffusion in our $KE$ scaling (see Equation \ref{eq:kappa_dep}), we suggest this is the dominant balance struck in our high-$Ra_F$ models.  This scaling appears as a common feature of high-$Ra$ convection in the laboratory setup (see e.g., Ahlers et al. 2009 and references therein).  Moreover, Boussinesq studies carried out using spherical geometry and symmetric, fixed temperature boundary conditions (in lieu of internal heating) have demonstrated a similar scaling relationship (Gastine et al. 2015).  

The scaling properties of our boundary layer hold interesting consequences for extrapolation to the Sun.  Perhaps it is most intuitive to think that the most realistic models of solar convection would be achieved by simulations with the same diffusivities or Rayleigh number as the Sun.  However, the radiative cooling which characterizes the upper boundary layer in the Sun differs markedly from the thermal conduction employed in our models.  As a result, extrapolating to the solar microscopic diffusivities may not actually reproduce the properties of solar convection. The convection in both the Sun and in our simulations is largely driven by intense buoyancy fluctuations within the thermal boundary layer. Since, the structure of that convection is likely tied to the boundary layer's width, it might be best to extrapolate our models to a common boundary-layer thickness. 

This exercise requires far less extrapolation in parameter space because the Sun's thermal boundary layer is only somewhat thinner than those achieved here.  Our most turbulent case possesses an $Ra_F$ of roughly $3\times10^7$ and a boundary layer width of about 3 Mm.  Were we to seek a boundary layer width of 1 Mm, consistent with the size of a typical solar granule, we would need to decrease $\kappa$ for that case by another factor of nine and increase $Ra_F$ by a factor of about 300.  While this would certainly be challenging to accomplish using modern computational resources, it is not impossible.  Alternatively, convective shells embodying a higher degree of density stratification might be examined to similarly achieve a thinner boundary layer.  In either case, fully spectral simulations such as these are, somewhat surprisingly, capable of resolving many of the relevant spatial scales in the solar convection zone.

If we were instead to follow our initial instinct and extrapolate our simulations to a Rayleigh number more characteristic of stellar convection zones (e.g., 10$^{20}$), the boundary layer would become unphysically thin. The resulting convection would likely be over-driven at small scales. It is thus possible that the convective velocity spectra achieved in these simulations is more similar to that occuring in an actual star than in a hypothetical ``stellar"-Rayleigh-number case.  This subtlety should be kept in mind when extrapolating results such as these to stellar-like parameter regimes.  Finally, we note that the measured prefactor for our scaling is almost certainly determined by the conductive physics at work in the boundary layer, and that the implementation of radiative cooling will likely modify the amplitude of those curves illustrated in Figures \ref{fig:ke_scaling}\textit{a},\textit{b}.

\subsection{Interaction with Rotation}
While our simulations were non-rotating, it is nonetheless interesting to speculate on how the behavior of the kinetic energy spectrum might impact the differential rotation.  The most reliable result from helioseismology with regard to the dynamics of the deep convection zone continues to be the Sun's internal-rotation profile obtained through global inversions of $p$-mode frequency splittings (e.g., Thompson et al. 2003).  Such inversions indicate that the surface differential rotation possesses a variation in angular velocity (decreasing by 30\% from equator to pole) that persists throughout the convection zone with little radial variation.  

Because of this robust observational constraint, the onset of anti-solar differential rotation in high $Ra_F$, rotating convection is troubling.  This behavior has been studied in a variety of stellar and astrophysical contexts, and it has been unambiguously linked to the degree of rotational influence felt by the convection (Gilman 1977; Glatzmaier \& Gilman 1982; Gastine et al. 2013; Guerrero et al. 2013; Gastine et al. 2014; K\"{a}pyl\"{a} et al. 2014; Featherstone \& Miesch 2015).  
The rotational constraint felt by the convection is characterized by the Rossby number, $Ro$, which expresses the ratio of a rotation period to a convective time scale.  It is typically defined as $Ro = u/2\Omega H$, where $u$ is a typical convective flow speed, $\Omega$ is the rotation rate of the star, and $H$ is some typical length scale (typically taken as the depth of the convective layer). 

A transition occurs between solar and anti-solar behavior when the convective and rotational timescales become comparable, and $Ro$ approaches unity.  Alternatively, one may define a local Rossby number $Ro_\ell = u_{rms}/\Omega\ell_c$, where $\ell_c$ is a characteristic length scale associated with the convection (not to be confused with the spherical harmonic degree $\ell\,$).   Such a definition is equally useful for characterizing the transition between solar and anti-solar states (Gastine et al. 2014).  The trend evident in the spectra of Figure \ref{fig:ke_spectra} suggests that $\ell_c$ will continue to decrease as $Ra_F$ is increased.  At the same time, our results indicate that $u_{rms}$ will remain fixed.  If this trend continues, $Ro_\ell$ will continue to increase as $Ra_F$ is increased, and anti-solar states, which naturally arise at high values of $Ro_\ell$, may become more prevalent.


We note that magnetism will almost certainly modify this discussion.  In fact, several studies indicate that the presence of a Lorentz force, which tends to inhibit the mixing of momentum by convection, increases the transitional value of $Ro$ in rotating systems (e.g., Fan \& Fang 2014; Karak et al. 2015; Simitev et al. 2015). How those magnetic effects relate to the spectral redistribution of energy discussed here remains to be examined.

\subsection{The Near-Surface Shear Layer}
Our results may also hold interesting consequences for the study of the solar near-surface shear layer.  That region of the convection zone, which spans the outer 5\% of the Sun by radius, is characterized by a diminishment of the angular velocity established within the bulk of the convection zone (e.g., Howe 2009).  Precisely how this region of shear is established is not yet completely understood, though it is likely determined by a transition from deep convection, which is thought to be rotationally-constrained, to near-surface convection, which is only weakly influenced by rotation (Miesch \& Hindman 2011; Hotta et al. 2014).    A number of numerical convection studies suggest that density stratification also plays an important role in the development of such a region of shear (K\"{a}pyl\"{a} et al. 2011; Gastine et al. 2013; Hotta et al. 2014).  One reason for this behavior may be the tendency of a background density stratification to accentuate differences in the speed of near-surface flows relative to deep flows, as shown in Figure \ref{fig:ke_radius}\textit{b}.  Higher values of $Ra_F$ also make this difference more pronounced (Figure \ref{fig:ke_radius}\textit{a}), though density stratification seems to be the dominant effect in the parameter space explored here.  Our results suggest that studies designed to systematically explore the development of a near-surface shear layer should focus on the high-$N_\rho$ and high-$Ra_F$ regime.

\subsection{Conclusions}
The results from this study indicate that care must be taken in the interpretation of simulated convective flows and, in particular, when carrying out comparisons between those flows and their solar/stellar counterparts.  The numerical results discussed here illustrate one means by which stellar convection simulations may naturally overestimate the low-wavenumber power in a convective velocity spectrum.  Namely, at sufficiently high Rayleigh number $Ra_F$, the integrated, dimensional kinetic energy saturates at a constant value.  That value is independent of the level of thermal diffusion.  At the same time, as diffusion is decreased and $Ra_F$ is increased, the convection naturally develops smaller-scale structure and a corresponding increase in high-wavenumber power.  As the integrated kinetic energy remains approximately constant, that high-wavenumber power must come at the expense of low-wavenumber power.  

Simulations that are not run with a sufficiently high value of $Ra_F$ will thus possess spectra that naturally disagree with the low-wavenumber measurements accessible to helioseismology.  This general trend may contribute to the disagreement between simulations and the deep time-distance measurements discussed in Hanasoge et al. (2012).  It is also possible that this trend may help explain the disagreement between photospheric convection simulations and the observed photospheric power spectra as discussed in Lord et al. (2014). While such photospheric convection models incorporate radiative-transfer effects that were not considered in this study, their convection is similarly constrained to transport a solar flux, and so it may develop a similar phenomenology.  Those simulations also employ open boundary conditions and Cartesian geometry, however, and thus a more thorough study of this effect needs to be carried out within that experimental setup before definitive conclusions may be drawn.

What is clear from this work is that stellar convection simulations must be run in the high-Rayleigh-number regime if they are to correctly capture even the grossest property of a star's convection, namely the integrated kinetic energy.  Unfortunately, even in the high-Rayleigh-number regime, it remains unclear how well such simulations can address the spectral distribution of that energy, particularly if they do not properly capture the scale of convective driving.  For the parameter space explored here, the spectral range $\ell \leq 10$ appears to represent a particularly sensitive region of the power spectrum.  

We conclude by noting that only relatively modest values of diffusion ($\kappa \leq 4\times10^{12}$ cm$^2$ s$^{-1}$) are required to reach the high-Rayleigh number regime indentified here.  However, virtually all stellar convection simulations incorporate rotation, and many now incorporate magnetism.  Our simulations incorporated neither.  The inclusion of those effects will almost certainly increase the critical Rayleigh number of the system (e.g., Chandrasekhar 1961), and so too the lower limit of the  high-Rayleigh-number regime.  It is thus timely to examine how magnetism and rotation modify these findings, and then assess where current models lie within the framework discussed here.  In the interim, the utility of detailed spectral comparisons between observations and simulations will remain questionable.

%% file: Appendices.tex
\appendix

\section{Algorithmic Accuracy Considerations}
The accuracy with which $Rayleigh$ solves the system of equations enumerated in \S\ref{sec:model} may be assessed by carrying out the benchmark exercises desribed in Jones et al. (2011).  We report on results from the steady hydrodynamic and magnetohydrodynamic benchmark exercises described in that work.  Those solutions each possess a convective pattern that is comprised of a single, sectoral spherical harmonic mode.  That pattern is temporally steady save for a uniform drift with respect to the rotating frame.  

The accuracy of a code may be checked by measuring the drift rate of the convective pattern, various components of the energy associated with that pattern, and by computing point-wise measurements of flow variables such as velocity, entropy, and magnetic field.  We provide a summary of the results obtained by \textit{Rayleigh} in Tables A.1 and A.2 below.  For reference, we also provide representative values from those reported in Jones et al. (2011).  Specifically, we provide the values achieved by the Glatzmaier code, along with relative differences between the Glatzmaier results and the $Rayleigh$ results.  \textit{Rayleigh's} results typically agree with the accepted benchmark results to within one tenth of one percent.

\joneshydro

\jonesmhd

\clearpage

\section{Summary of Model Parameters and Results}
In the four tables that follow, we report on the input and output parameters from our set of 63 simulations.  A separate table of parameters is presented for each value of $N_\rho$ used in this study.  Each simulation is specified by three physical input parameters: the thermal diffusivity $\kappa$, the luminosity of the simulation $L_\star$, and the bulk flux Rayleigh number $Ra_F$ as defined by Equation (\ref{eq:ranum}). We report the thermal diffusivity $\kappa$ in units of 10$^{12}$ cm$^2$ s$^{-1}$, denoting it by $\kappa_{12}$ in these tables.  Similarly, we report luminosities that have been normalized by the solar luminosity $L_\odot=3.846\times10^{33}$ erg s$^{-1}$.  Additionally, we report on the resolution used in each simulation, listing the maximum spherical harmonic and Chebyshev polynomial degrees employed ($\ell_{max}$ and $n_{max}$ respectively).  All variables have been de-aliased using the 2/3 rule.  The number of collocation points in radius $N_R$ is given by $N_R = \frac{3}{2}(n_{max}+1)$.  Similarly, the number of collocation points in theta $N_\theta$ is given by $N_\theta = \frac{3}{2}(\ell_{max}+1)$.  The number of azimuthal points $N_\phi$ is always twice $N_\theta$. 

We also report on the measured outputs for each simulation.  These are the dimensional kinetic energy $KE$, the non-dimensional kinetic energy $\widehat{KE}$, the relative velocity difference $\delta V$, the fractional convective flux $f_{conv}$, and the boundary layer width $w_{BL}$.  Definitions for these values are provided in Equations (\ref{ke_def}), (\ref{eq:kend}), (\ref{eq:delv}), (\ref{eq:fconv}), and (\ref{eq:width}) respectively.  Finally, while not discussed in this paper, we report on two Reynolds numbers measured for each simulation.  The first is the bulk Reynolds number, which we denote by $Re$ and define as
\begin{equation}
\label{eq:reynolds}
Re = \frac{\sqrt{\widetilde{|\vec{v}|^2}}H}{\nu},
\end{equation} 
where $H$ and the tilde retain the same meaning as in the rest of this paper, representing the shell depth and a volume average respectively.  The density stratification employed in these simulations means that the velocity amplitude varies substantially throughout the shell.  As such, we also report the peak Reynolds number $Re_{peak}$, which we define as
\begin{equation}
\label{eq:reynolds_peak}
Re_{peak} = \frac{\mathrm{max}\, v_{rms}(r) H}{\nu},
\end{equation}
where $v_{rms}$ is the $rms$ velocity amplitude calculated at each depth.  Both measures of the velocity were averaged in time over 1,000 days.

\begin{table}[p]
\label{mhd_table}\centering\small
\begin{tabular}[t]{rrrrr|rrrrrrr}\\
\multicolumn{12}{c}{Table B.1:  $N_\rho=1$ Simulation Parameters}\\\hline
\multicolumn{5}{c}{Input Parameters} &  \multicolumn{7}{c}{Measured Output}\\\hline
\multicolumn{1}{c}{$\kappa_{12}$} & \multicolumn{1}{c}{$L_\star/L_\odot$} & \multicolumn{1}{c}{$Ra_F$} &\multicolumn{1}{c}{$n_{max}$} &\multicolumn{1}{c}{$\ell_{max}$}&\multicolumn{1}{c}{$KE$}&\multicolumn{1}{c}{$\widehat{KE}$}&\multicolumn{1}{c}{$\delta V$} &\multicolumn{1}{c}{$f_{conv}$}&\multicolumn{1}{c}{$w_{BL}$} &\multicolumn{1}{c}{$Re$} & \multicolumn{1}{c}{$Re_{peak}$}\\ 
\multicolumn{1}{c}{~} & ~ & ~ & ~ & ~ &\multicolumn{1}{c}{(10$^{38}$ erg)} & ~ & ~ & ~ &\multicolumn{1}{c}{(Mm)} & ~ & ~
\\\hline\hline
\input 1.tex
\hline

\end{tabular}
\end{table}

\begin{table}[t]
\label{mhd_table}\centering\small
\begin{tabular}[t]{rrrrr|rrrrrrr}\\
\multicolumn{12}{c}{Table B.2:  $N_\rho=2$ Simulation Parameters}\\\hline
\multicolumn{5}{c}{Input Parameters} &  \multicolumn{7}{c}{Measured Output}\\\hline
\multicolumn{1}{c}{$\kappa_{12}$} & \multicolumn{1}{c}{$L_\star/L_\odot$} & \multicolumn{1}{c}{$Ra_F$} &\multicolumn{1}{c}{$n_{max}$} &\multicolumn{1}{c}{$\ell_{max}$}&\multicolumn{1}{c}{$KE$}&\multicolumn{1}{c}{$\widehat{KE}$}&\multicolumn{1}{c}{$\delta V$} &\multicolumn{1}{c}{$f_{conv}$}&\multicolumn{1}{c}{$w_{BL}$} &\multicolumn{1}{c}{$Re$} & \multicolumn{1}{c}{$Re_{peak}$}\\ 
\multicolumn{1}{c}{~} & ~ & ~ & ~ & ~ &\multicolumn{1}{c}{(10$^{38}$ erg)} & ~ & ~ & ~ &\multicolumn{1}{c}{(Mm)} & ~ & ~
\\\hline\hline
\input 2.tex
\hline

\end{tabular}
\end{table}

\begin{table}[t]
\label{mhd_table}\centering\small
\begin{tabular}[t]{rrrrr|rrrrrrr}\\
\multicolumn{12}{c}{Table B.3:  $N_\rho=3$ Simulation Parameters}\\\hline
\multicolumn{5}{c}{Input Parameters} &  \multicolumn{7}{c}{Measured Output}\\\hline
\multicolumn{1}{c}{$\kappa_{12}$} & \multicolumn{1}{c}{$L_\star/L_\odot$} & \multicolumn{1}{c}{$Ra_F$} &\multicolumn{1}{c}{$n_{max}$} &\multicolumn{1}{c}{$\ell_{max}$}&\multicolumn{1}{c}{$KE$}&\multicolumn{1}{c}{$\widehat{KE}$}&\multicolumn{1}{c}{$\delta V$} &\multicolumn{1}{c}{$f_{conv}$}&\multicolumn{1}{c}{$w_{BL}$} &\multicolumn{1}{c}{$Re$} & \multicolumn{1}{c}{$Re_{peak}$}\\ 
\multicolumn{1}{c}{~} & ~ & ~ & ~ & ~ &\multicolumn{1}{c}{(10$^{38}$ erg)} & ~ & ~ & ~ &\multicolumn{1}{c}{(Mm)} & ~ & ~
\\\hline\hline
\input 3.tex
\hline

\end{tabular}
\end{table}

\begin{table}[t]
\label{mhd_table}\centering\small
\begin{tabular}[t]{rrrrr|rrrrrrr}\\
\multicolumn{12}{c}{Table B.4:  $N_\rho=4$ Simulation Parameters}\\\hline
\multicolumn{5}{c}{Input Parameters} &  \multicolumn{7}{c}{Measured Output}\\\hline
\multicolumn{1}{c}{$\kappa_{12}$} & \multicolumn{1}{c}{$L_\star/L_\odot$} & \multicolumn{1}{c}{$Ra_F$} &\multicolumn{1}{c}{$n_{max}$} &\multicolumn{1}{c}{$\ell_{max}$}&\multicolumn{1}{c}{$KE$}&\multicolumn{1}{c}{$\widehat{KE}$}&\multicolumn{1}{c}{$\delta V$} &\multicolumn{1}{c}{$f_{conv}$}&\multicolumn{1}{c}{$w_{BL}$} &\multicolumn{1}{c}{$Re$} & \multicolumn{1}{c}{$Re_{peak}$}\\ 
\multicolumn{1}{c}{~} & ~ & ~ & ~ & ~ &\multicolumn{1}{c}{(10$^{38}$ erg)} & ~ & ~ & ~ &\multicolumn{1}{c}{(Mm)} & ~ & ~
\\\hline\hline
\input 4.tex
\hline

\end{tabular}
\end{table}

\clearpage

\section{Polytropic Background State}
For these experiments, we employ a polytropic background state as formulated in the anelastic benchmark paper of Jones et al. (2011).
We assume that the gravitational acceleration $g(r)$ varies as $GM_{i}/r^2$
within the shell, where $M_{i}$ is the mass interior to the base of the
convection zone and $G$ is the gravitational constant.  The anelastic equations then admit
an adiabatically stratified, polytropic atmosphere as the reference
state:
\begin{equation}
\label{polytrope}
\rho_0 = \rho_i\left(\frac{\zeta}{\zeta_i}\right)^n,~~~~T_0=T_i\frac{\zeta}{\zeta_i},~~~~p_0=p_i\left(\frac{\zeta}{\zeta_i}\right)^{n+1},
\end{equation}
where the subscript ``\textit{i}" denotes the value of a quantity at the inner boundary,
and \textit{n} is the polytropic index.  The radial variation of the reference state
is captured by the variable $\zeta$, defined as
\begin{equation}
\zeta=c_0+\frac{c_1H}{r},
\end{equation}
where $H = r_o-r_i$ is the depth of the convection zone.  The constants $c_0$ and $c_1$ are given by
\begin{equation}
c_0=\frac{2\zeta_o-\beta-1}{1-\beta},~~~~c_1=\frac{(1+\beta)(1-\zeta_o)}{(1-\beta)^2},
\end{equation}
with
\begin{equation}
\zeta_o=\frac{\beta+1}{\beta\mathrm{exp}(N_\rho/n)+1},~~~~\zeta_i=\frac{1+\beta-\zeta_o}{\beta}.
\end{equation}
Here $\zeta_i$ and $\zeta_o$ are the values of $\zeta$ on the inner and outer
boundaries, $\beta=r_i/r_o$, and $N_\rho$ is the number of density scale heights
across the shell.

%% file: 1.tex
0.5 &  1.00 & $3.24\times10^{7}$ &  85 & 1023 & 26.99 & 36,649.8 & 0.082 & 0.934 &  6.47 & 271.7 & 355.1\\
  1 &  1.00 & $4.05\times10^{6}$ &  85 & 511 & 25.41 &  8,626.2 & 0.056 & 0.894 &  9.32 & 130.5 & 167.0\\
  2 &  1.00 & $5.07\times10^{5}$ &  85 & 255 & 23.86 &  2,025.1 & 0.027 & 0.830 & 14.09 &  63.3 &  77.7\\
  4 &  1.00 & $6.33\times10^{4}$ &  85 & 255 & 19.82 &   420.5 & 0.013 & 0.725 & 20.00 &  29.0 &  36.3\\
  8 &  1.00 & $7.92\times10^{3}$ &  85 & 255 & 10.66 &    56.5 & 0.012 & 0.533 & 26.02 &  10.8 &  15.2\\
  9 &  1.00 & $5.56\times10^{3}$ &  85 & 127 &  8.77 &    36.7 & 0.013 & 0.469 & 29.13 &   8.7 &  12.3\\
 10 &  1.00 & $4.05\times10^{3}$ &  85 & 127 &  6.95 &    23.6 & 0.015 & 0.379 & 33.91 &   6.9 &   9.9\\
 11 &  1.00 & $3.05\times10^{3}$ &  85 & 127 &  5.12 &    14.4 & 0.013 & 0.345 & 35.35 &   5.4 &   7.9\\
 12 &  1.00 & $2.35\times10^{3}$ &  85 & 127 &  3.62 &     8.5 & 0.014 & 0.249 & 39.42 &   4.2 &   6.2\\
 13 &  1.00 & $1.84\times10^{3}$ &  85 & 127 &  2.21 &     4.4 & 0.015 & 0.166 & 42.37 &   3.0 &   4.3\\

%% file: 2.tex
  1 &  1.00 & $1.23\times10^{7}$ &  85 & 511 & 34.67 & 16,095.4 & 0.416 & 0.932 &  6.80 & 189.4 & 243.2\\
  2 &  1.00 & $1.54\times10^{6}$ &  85 & 255 & 33.08 &  3,839.6 & 0.310 & 0.887 &  9.66 &  91.9 & 120.0\\
  4 &  1.00 & $1.93\times10^{5}$ &  85 & 255 & 30.92 &   897.0 & 0.179 & 0.807 & 13.86 &  44.8 &  60.8\\
  8 &  1.00 & $2.41\times10^{4}$ &  85 & 255 & 23.94 &   173.6 & 0.084 & 0.689 & 19.65 &  20.5 &  30.1\\
  9 &  1.00 & $1.69\times10^{4}$ &  85 & 85 & 20.28 &   116.3 & 0.091 & 0.644 & 19.70 &  16.7 &  24.2\\
 10 &  1.00 & $1.23\times10^{4}$ &  85 & 85 & 19.41 &    90.1 & 0.073 & 0.620 & 21.62 &  15.0 &  23.5\\
 11 &  1.00 & $9.28\times10^{3}$ &  85 & 85 & 17.25 &    66.2 & 0.102 & 0.565 & 21.62 &  12.5 &  18.5\\
 12 &  1.00 & $7.14\times10^{3}$ &  85 & 85 & 15.53 &    50.1 & 0.099 & 0.526 & 22.72 &  10.9 &  16.3\\
 13 &  1.00 & $5.62\times10^{3}$ &  85 & 85 & 12.71 &    34.9 & 0.087 & 0.500 & 24.15 &   9.3 &  15.3\\
 14 &  1.00 & $4.50\times10^{3}$ &  85 & 85 & 10.84 &    25.7 & 0.099 & 0.422 & 25.97 &   7.8 &  12.4\\
 15 &  1.00 & $3.66\times10^{3}$ &  85 & 85 &  8.99 &    18.6 & 0.100 & 0.382 & 27.17 &   6.9 &  10.9\\
 16 &  1.00 & $3.01\times10^{3}$ &  85 & 85 &  7.06 &    12.8 & 0.106 & 0.330 & 28.79 &   5.5 &   9.3\\
 17 &  1.00 & $2.51\times10^{3}$ &  85 & 85 &  5.64 &     9.1 & 0.118 & 0.290 & 29.75 &   4.7 &   8.2\\
 18 &  1.00 & $2.12\times10^{3}$ &  85 & 85 &  3.33 &     4.8 & 0.115 & 0.194 & 32.23 &   3.5 &   5.9\\
 19 &  1.00 & $1.80\times10^{3}$ &  85 & 85 &  2.36 &     3.0 & 0.111 & 0.139 & 33.50 &   2.7 &   4.6\\
 20 &  1.00 & $1.54\times10^{3}$ &  85 & 85 &  1.16 &     1.3 & 0.104 & 0.074 & 34.89 &   1.8 &   3.1\\

%% file: 3.tex
  1 &  1.00 & $2.14\times10^{7}$ &  85 & 1023 & 36.22 & 20,057.7 & 1.323 & 0.948 &  4.96 & 229.8 & 320.4\\
  2 &  1.00 & $2.68\times10^{6}$ &  85 & 511 & 34.73 &  4,807.6 & 1.094 & 0.912 &  6.93 & 114.1 & 166.5\\
  4 &  1.00 & $3.35\times10^{5}$ &  85 & 263 & 33.42 &  1,156.7 & 0.799 & 0.847 &  9.83 &  56.3 &  85.6\\
  6 &  1.00 & $9.93\times10^{4}$ &  85 & 127 & 32.92 &   506.4 & 0.617 & 0.791 & 12.18 &  37.5 &  58.2\\
  8 &  1.00 & $4.19\times10^{4}$ &  85 & 127 & 31.62 &   273.5 & 0.452 & 0.736 & 14.09 &  27.5 &  42.4\\
 12 &  1.00 & $1.24\times10^{4}$ &  85 & 127 & 22.08 &    84.9 & 0.275 & 0.623 & 16.03 &  15.4 &  24.0\\
 16 &  1.00 & $5.23\times10^{3}$ &  42 & 127 & 12.02 &    26.0 & 0.266 & 0.442 & 19.06 &   9.0 &  15.6\\
 24 &  1.00 & $1.55\times10^{3}$ &  42 & 63 &  2.05 &     2.0 & 0.198 & 0.122 & 23.77 &   2.5 &   4.5\\
0.5 &  0.19 & $3.25\times10^{7}$ & 170 & 85 & 12.27 & 27,173.6 & 1.353 & 0.953 &  4.65 & 265.3 & 363.5\\
  1 &  0.29 & $6.16\times10^{6}$ &  85 & 1023 & 15.37 &  8,512.9 & 1.189 & 0.929 &  6.03 & 151.4 & 215.8\\
  2 &  0.44 & $1.17\times10^{6}$ &  85 & 511 & 19.68 &  2,723.7 & 1.011 & 0.890 &  7.93 &  86.5 & 129.8\\
  4 &  0.66 & $2.21\times10^{5}$ &  85 & 263 & 25.39 &   878.7 & 0.710 & 0.830 & 10.60 &  49.0 &  74.4\\
  6 &  0.84 & $8.35\times10^{4}$ &  85 & 127 & 28.96 &   445.4 & 0.562 & 0.781 & 12.70 &  35.3 &  55.0\\
  1 &  0.12 & $2.68\times10^{6}$ &  85 & 1023 &  8.71 &  4,820.8 & 1.096 & 0.912 &  6.94 & 113.8 & 166.1\\
  2 &  0.25 & $6.70\times10^{5}$ &  85 & 511 & 13.44 &  1,860.7 & 0.898 & 0.873 &  8.75 &  71.2 & 105.9\\
  4 &  0.50 & $1.68\times10^{5}$ &  85 & 263 & 21.00 &   726.7 & 0.713 & 0.815 & 11.09 &  44.3 &  68.5\\
  6 &  0.75 & $7.44\times10^{4}$ &  85 & 127 & 27.12 &   417.1 & 0.545 & 0.773 & 12.85 &  34.2 &  52.7\\

%% file: 4.tex
  1 &  1.00 & $2.84\times10^{7}$ & 170 & 1023 & 34.38 & 20,892.4 & 3.216 & 0.958 &  3.37 & 254.0 & 388.4\\
  2 &  1.00 & $3.55\times10^{6}$ &  85 & 511 & 32.21 &  4,892.5 & 2.691 & 0.925 &  4.66 & 124.2 & 193.0\\
  4 &  1.00 & $4.43\times10^{5}$ &  85 & 511 & 30.91 &  1,173.8 & 1.961 & 0.864 &  6.63 &  61.5 &  98.3\\
  8 &  1.00 & $5.54\times10^{4}$ &  85 & 255 & 29.49 &   280.0 & 1.145 & 0.755 &  9.42 &  29.9 &  48.5\\
  9 &  1.00 & $3.89\times10^{4}$ &  85 & 127 & 29.04 &   217.9 & 1.060 & 0.732 &  9.99 &  26.5 &  43.5\\
 10 &  1.00 & $2.84\times10^{4}$ &  85 & 127 & 28.72 &   174.5 & 0.901 & 0.712 & 10.54 &  23.5 &  38.3\\
 11 &  1.00 & $2.13\times10^{4}$ &  85 & 127 & 27.03 &   135.7 & 0.800 & 0.688 & 11.00 &  20.9 &  34.2\\
 12 &  1.00 & $1.64\times10^{4}$ &  85 & 127 & 25.64 &   108.2 & 0.697 & 0.661 & 11.47 &  18.8 &  31.1\\
 13 &  1.00 & $1.29\times10^{4}$ &  85 & 127 & 22.53 &    81.0 & 0.479 & 0.624 & 11.99 &  16.4 &  27.6\\
 14 &  1.00 & $1.03\times10^{4}$ &  85 & 127 & 20.44 &    63.4 & 0.410 & 0.594 & 12.32 &  14.6 &  24.9\\
 15 &  1.00 & $8.41\times10^{3}$ &  85 & 127 & 18.45 &    49.8 & 0.409 & 0.557 & 12.57 &  13.0 &  22.2\\
 16 &  1.00 & $6.93\times10^{3}$ &  85 & 127 & 16.32 &    38.7 & 0.383 & 0.522 & 12.81 &  11.6 &  20.0\\
 17 &  1.00 & $5.78\times10^{3}$ &  85 & 85 & 14.18 &    29.8 & 0.380 & 0.486 & 13.04 &  10.2 &  18.0\\
 18 &  1.00 & $4.87\times10^{3}$ &  85 & 85 & 12.38 &    23.2 & 0.410 & 0.436 & 13.37 &   9.2 &  16.3\\
 19 &  1.00 & $4.14\times10^{3}$ &  85 & 85 & 10.40 &    17.5 & 0.379 & 0.393 & 13.62 &   7.9 &  14.3\\
 20 &  1.00 & $3.55\times10^{3}$ &  85 & 85 &  7.21 &    10.9 & 0.177 & 0.336 & 14.09 &   6.4 &  12.5\\
 21 &  1.00 & $3.06\times10^{3}$ &  85 & 85 &  5.49 &     7.6 & 0.186 & 0.270 & 14.50 &   5.3 &  10.4\\
 22 &  1.00 & $2.67\times10^{3}$ &  85 & 85 &  3.69 &     4.6 & 0.196 & 0.196 & 15.01 &   4.2 &   8.3\\
 23 &  1.00 & $2.33\times10^{3}$ &  85 & 85 &  3.10 &     3.6 & 0.498 & 0.180 & 15.03 &   3.8 &   7.2\\
 24 &  1.00 & $2.05\times10^{3}$ &  85 & 85 &  2.96 &     3.1 & 0.475 & 0.166 & 15.15 &   3.4 &   6.5\\